\documentclass[prl,superscriptaddress,showpacs,twocolumn,a4paper]{revtex4}
\usepackage{graphicx}

\newcommand{\centreline}{\centerline}

\newcommand{\TiO}{TiO$_{2}$}
\newcommand{\TiCoO}{Ti$_{1-x}$Co$_{x}$O$_{2-\delta}$}

\newcommand{\etal}{\textit{et al.}}

\begin{document}

\preprint{Version 3}

\title{Signature of Carrier-Induced Ferromagnetism in \TiCoO : \\
       Exchange Interaction Between High-Spin Co$^{2+}$ and the
       Ti~$3d$ Conduction Band}

\author{J.~W.~Quilty}
\affiliation{PRESTO, Japan Science and Technology Corporation (JST),
             4-1-8 Honcho Kawaguchi, Saitama 332-0012, Japan}
\author{A.~Shibata}
\affiliation{Department of Complexity Science and Engineering,
             University of Tokyo, 5-1-5 Kashiwanoha, Chiba 277-8581,
             Japan}
\author{J.-Y.~Son}
\affiliation{PRESTO, Japan Science and Technology Corporation (JST),
             4-1-8 Honcho Kawaguchi, Saitama 332-0012, Japan}
\author{K.~Takubo}
\affiliation{Department of Physics, University of Tokyo, Hongo 7-3-1,
             Bunkyo-ku, Tokyo 113-0033, Japan}
\author{T.~Mizokawa}
\affiliation{PRESTO, Japan Science and Technology Corporation (JST),
             4-1-8 Honcho Kawaguchi, Saitama 332-0012, Japan}
\affiliation{Department of Physics, University of Tokyo, Hongo 7-3-1,
             Bunkyo-ku, Tokyo 113-0033, Japan}
\affiliation{Department of Complexity Science and Engineering,
             University of Tokyo, 5-1-5 Kashiwanoha, Chiba 277-8581,
             Japan}
\author{H.~Toyosaki}
\author{T.~Fukumura}
\affiliation{Institute for Materials Research, Tohoku University,
             Sendai 980-8577, Japan}
\author{M.~Kawasaki}
\affiliation{Institute for Materials Research, Tohoku University,
             Sendai 980-8577, Japan}
\affiliation{Combinatorial Materials Exploration and Technology,
             Tsukuba 305-0044, Japan}

\begin{abstract}

X-ray photoemission spectroscopy measurements were performed on
thin-film samples of rutile \TiCoO\ to reveal the electronic
structure. The Co 2$p$ core level spectra indicate that the Co ions
take the high-spin Co$^{2+}$ configuration, consistent with
substitution on the Ti site. The high spin state and the shift due to
the exchange splitting of the conduction band suggest strong
hybridization between carriers in the Ti~$3d$ $t_{2g}$ band and the
$t_{2g}$ states of the high-spin Co$^{2+}$. These observations support
the argument that room temperature ferromagnetism in \TiCoO\ is
intrinsic.

\end{abstract} 


\maketitle

Reports of ferromagnetism at room temperature in anatase and rutile
\TiCoO\ \cite{Matsumoto2001:sci291,Matsumoto2001:jjap40} have
attracted a considerable amount of attention
\cite{Toyosaki2004:NatMat3,Park2002:prb65,Geng2003:prb68,Kim2003:prl90,Shinde2003:prb67,Shinde2004:prl92}.
If an intrinsic effect, this ubiquitous semiconductor would join the
class of diluted magnetic semiconductors (DMS) which include colossal
magnetoresistance and the anomalous hall effect among their physical
properties. That the maximum Curie temperature of known DMS materials
are around 160~K \cite{Chiba2003:apl82} makes reports of room
temperature ferromagnetism in a semiconductor particularly noteworthy.
It has proved difficult, however, to determine whether the observed
ferromagnetism is an intrinsic effect due solely to the substitution
of Co on the Ti sites or an extrinsic effect due to the presence of
clustered metallic Co.

Early reports of room temperature ferromagnetism in both anatase and
rutile \TiCoO\ were made by Matsumoto \etal\
\cite{Matsumoto2001:sci291,Matsumoto2001:jjap40}. The study of anatase
combinatorial libraries \cite{Matsumoto2001:sci291} concluded that the
Co ions occupy the Ti cation sites and that \TiCoO\ can be viewed as a
diluted magnetic (wide-gap) semiconductor. On the other hand, it has
been claimed that the ferromagnetism is attributable to Co
nanoclusters in the \TiCoO\ thin films rather than the substituted Co
ions \cite{Kim2003:prl90,Shinde2004:prl92}. The origin of
ferromagnetism in both anatase and rutile \TiCoO\ remains the subject
of intense debate.

Very recently, anomalous Hall effect measurements on systematically
produced rutile \TiCoO\ films with different carrier doping levels
\cite{Toyosaki2004:NatMat3} have confirmed room temperature
ferromagnetism in this material, strongly suggesting that the
ferromagnetism is intrinsic and not the result of Co clustering. At
the same time as the anomalous Hall effect measurements were reported
a magnetization study of highly reduced rutile thin film samples
revealed the co-occurrence of superparamagnetism and anomalous Hall
effect, casting doubt on the reliability of the anomalous Hall effect
for determining whether ferromagnetism in \TiCoO\ is intrinsic or
extrinsic \cite{Shinde2004:prl92}. Determination of the effects of the
Co impurity on the electronic structure and the electronic
configuration of the Co impurity in \TiCoO\ is thus very important for
understanding the origins of the room temperature ferromagnetism in
this potential DMS.

This letter reports the results of x-ray photoemission spectroscopy
(XPS) measurements of the electronic structure of rutile phase \TiCoO\
thin films. These results show Co ions uniformly substituted on the Ti
cation sites without detectable clustering, supporting the idea that
the observed room temperature ferromagnetism is an intrinsic property
of these films.

Rutile \TiCoO\ thin films were taken from the same batches as the
samples studied in Ref.~\onlinecite{Toyosaki2004:NatMat3}, where room
temperature ferromagnetism was observed. These films were deposited by
laser molecular beam epitaxy on TiO$_{2}$-buffered sapphire
substrates, with $\delta$ controlled by varying the oxygen partial
pressure from 10$^{-4}$--10$^{-8}$ torr \cite{Toyosaki2004:NatMat3}.
This contrasts with the highly reduced films grown in high vacuum
which show Co clustering \cite{Shinde2004:prl92}. Further details of
sample fabrication may be found in
Ref.~\onlinecite{Toyosaki2004:NatMat3}. A JEOL JPS-9200 spectrometer,
with the Al K$\alpha$ line (h$\nu$= 1486.6~eV) as the x-ray source,
was used to measure XPS spectra with an energy resolution of 0.6~eV
under a base pressure of the order of $1\times10^{-10}$~Torr. Binding
energies were calibrated to the Au~$4f$ core level peaks of a gold
reference sample and the intensity of all spectra shown has been
normalised by peak height unless otherwise stated. To remove the
effects of band bending a Nd:YAG pulsed laser was used at an energy of
3.5~eV (355 nm) with frequency 30~Hz. The maximum incident energy
density at the sample surface used was 200~$\mu$J/cm$^{2}$ per pulse
and the effective pulse width was 6~ns.

\begin{figure}
    \centreline{\includegraphics{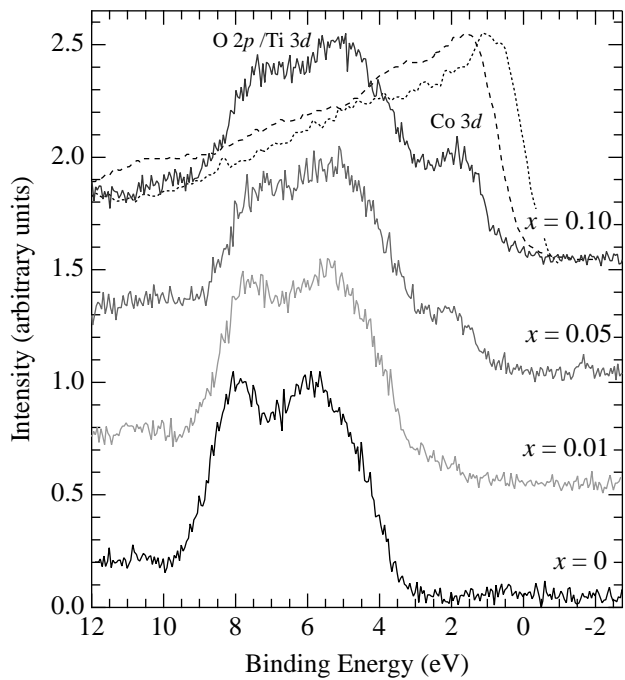}}
    \caption{Valence band x-ray photoemission spectra as a function of
             Co concentration $x$. Included for comparison are valence
             band spectra of metallic Co (dotted line) and CoO (dashed
             line) from Refs.~[\onlinecite{vanElp1991:prb44},
             \onlinecite{Galtayries1999:jesrp98}].}
    \label{fig:valence_xDep}
\end{figure}

\begin{figure}
    \centreline{\includegraphics{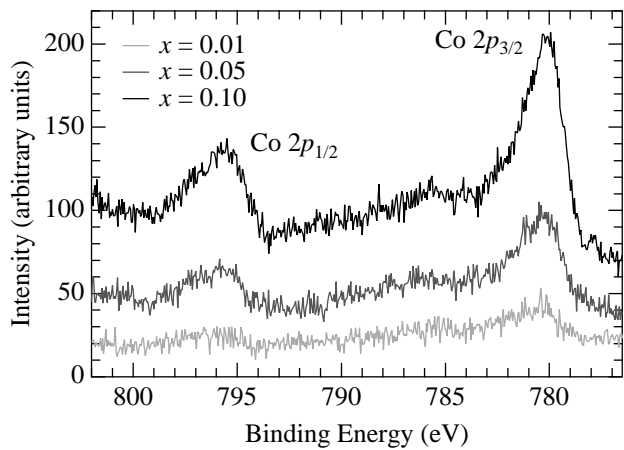}}
    \caption{Co~$2p$ core level x-ray photoemission spectra as a
             function of Co concentration $x$.}
    \label{fig:Co2p_xDep}
\end{figure}

\TiCoO\ valence band spectra for $x = 0$, 0.01, 0.05 and 0.10 are
shown in Fig.~\ref{fig:valence_xDep} and corresponding Co~$2p$ spectra
are presented in Fig.~\ref{fig:Co2p_xDep}. The pure \TiO\ valence band
spectrum shows a band gap extending 3~eV below the Fermi level to the
valence states which are derived from O~$2p$ and Ti~$3d$
\cite{Woicik2002:prl89}. Co substitution shifts the valence band to
lower binding energy by up to 0.5~eV at $x = 0.10$ and causes a broad
peak to appear in the spectra at around 2~eV, the intensity of which
increases with Co concentration. The Co~$2p$ core level spectra
exhibit a main Co~$2p_{3/2}$ peak at 780.1~eV, accompanied by a
relatively weak satellite peak around 786~eV, with the Co~$2p_{1/2}$
peak appearing at 795.6~eV ($\Delta E = 15.5$~eV). There is little
variation of the spectral lineshapes and peak binding energies with Co
concentration, showing that the local electronic structure of Co
impurities in \TiCoO\ is largely independent of $x$.
\nocite{XPSHandbook1992,Cattaruzza1998:apl73,Wang2004:jap95,Endnote_condmat0507193}
\nocite{Furdyna1982:jap53,vanElp1991:prb44,Saitoh1997:prb56,Bocquet1992:prb46}
\nocite{Potze1995:prb51,Weng2004:prb69,Yao2004:jap95,Galtayries1999:jesrp98}

Both the Co~$2p$ and valence band spectra show the signature of
well-substituted Co ions at the film surface. First, the Co~$2p_{3/2}$
binding energy, Co~$2p_{3/2}$--$2p_{1/2}$ separation and satellite
structure of the \TiCoO\ spectra are significantly different from
those of metallic Co and Co nanoclusters
\cite{XPSHandbook1992,Wang2004:jap95,Cattaruzza1998:apl73} which show
Co~$2p_{3/2}$ at 778.3--778.5~eV, $\Delta E = 15.0$~eV and lack any
satellite structure. Second, there is no discernible clustered
metallic Co component in either the valence band (cf.\ the dotted line
in Fig.~\ref{fig:valence_xDep}) or Co~$2p$ spectra, to an estimated
detection threshold of 0.5 at.~\%. For the observed moment of just
over 1~$\mu_{B}/$Co in the $x = 0.05$ and 0.10 films to be due to
metallic Co clusters would require 65\% of the total Co to cluster
\cite{Endnote_condmat0507193}. Such a large amount of clustered
cobalt should be clearly visible in Figs.~\ref{fig:valence_xDep} and
\ref{fig:Co2p_xDep}. Nevertheless, the possibility of a
well-substituted surface with metallic Co clusters segregated to the
bulk of the film can not be categorically excluded by a
surface-sensitive technique like XPS. The possibility of CoO at the
film surface is excluded by the large differences in the valence band
(cf.\ the dashed line in Fig.~\ref{fig:valence_xDep}) and Co~$2p$ line
shapes. In particular, while the Co~$2p$ main peak of CoO is wide due
to lattice effects, that of \TiCoO\ is rather narrow because the Co
ions are isolated in \TiCoO.

The feature at 2~eV in Fig.~\ref{fig:valence_xDep} is indicative of
the formation of an impurity band derived from the Co~$3d$ orbitals
and this observation implies that the donor level of the Co impurity
is located just above the valence band maximum of \TiO\ in the dilute
limit. These donor levels grow with Co doping and finally form the
Co~$3d$ impurity band. Although it is impossible to see the acceptor
level using photoemission spectroscopy, the energy difference between
the donor and acceptor levels is approximately given by the charge
transfer energy obtained from cluster model analysis of the Co 2$p$
spectra. The Co impurity band is located 2~eV below the Fermi level
for $x = 0.10$ and the entire band is shifted to lower binding energy
by around 0.5~eV, as estimated later from Ti~$2p$ and O~$1s$ core
level spectra. The donor level is thus located 2.5~eV below the Fermi
level in the dilute limit. As this value is comparable to the
charge-transfer energy of 4~eV, it is expected that the acceptor level
of the Co impurity is almost degenerate in energy with the conduction
band minimum. In addition, since the TiO$_6$ octahedra are edge
sharing in the rutile structure, direct hopping between the
substituted Co~3$d$ $t_{2g}$ orbital and the neighbouring Ti~3$d$
$t_{2g}$ orbital is expected to be large. These effects will
enhance the exchange coupling between the Co 3$d$ spins and the
carriers in the Ti 3$d$ conduction band \cite{Furdyna1982:jap53}.

\begin{figure}
    \centreline{\includegraphics{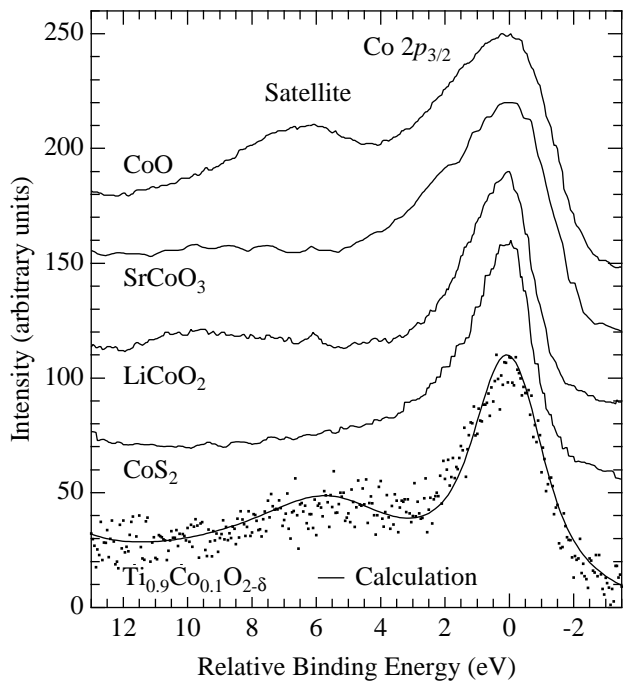}}
    \caption{Comparison of the Co~$2p$ spectra from 
             CoO \cite{vanElp1991:prb44}, 
             SrCoO$_{3}$ \cite{Saitoh1997:prb56}, 
             LiCoO$_{2}$ \cite{vanElp1991:prb44} and
             CoS$_{2}$ \cite{Bocquet1992:prb46} (solid lines) with
             that of Ti$_{0.9}$Co$_{0.1}$O$_{2-\delta}$ (dots).
             The solid line through the Ti$_{0.9}$Co$_{0.1}$O$_{2-\delta}$
             spectrum shows the result of a cluster model calculation 
             for CoO$_{6}$.}
    \label{fig:Co2p_Comparison}
\end{figure}

Interstitial dopant substitution is a possibility in the rutile
structure and to determine whether Co substitutes on the Ti site,
rather than interstitially, it is necessary to determine the
electronic state of the Co ion. Figure~\ref{fig:Co2p_Comparison} shows
a comparison of the measured Co~$2p$ spectrum for $x = 0.10$ (dots)
with the results of cluster model calculations (the solid line
superimposed on the experimental data). An integral background has been
added to the calculated spectrum. The cluster model calculations were
performed for CoO$_{6}$ clusters and the best agreement with the
measured spectrum was found for Co in the high spin Co$^{2+}$ state,
with parameters $\Delta = 4.0$~eV, $U = 6.5$~eV and $(pd\sigma) =
-1.1$~eV. As interstitially substituted Co is predicted to take the
low spin state \cite{Geng2003:prb68}, the observation of a high spin
Co$^{2+}$ state in the films indicates that the Co atoms are
predominantly substituted on the Ti site.

Comparison with XPS spectra obtained from CoO, LiCoO$_{2}$, CoS$_{2}$
and SrCoO$_{3}$
\cite{vanElp1991:prb44,Bocquet1992:prb46,Saitoh1997:prb56,Potze1995:prb51},
which exhibit a variety of Co electronic states, is also helpful.
These spectra, with binding energies normalised at the Co~$2p_{3/2}$
peak, are included in Fig.~\ref{fig:Co2p_Comparison} and show high
spin Co$^{2+}$ (CoO \cite{vanElp1991:prb44}), low spin Co$^{3+}$
(LiCoO$_{2}$ \cite{vanElp1991:prb44}), low spin Co$^{2+}$ (CoS$_{2}$
\cite{Bocquet1992:prb46}) and low or intermediate spin Co$^{4+}$
(SrCoO$_{3}$ \cite{Saitoh1997:prb56,Potze1995:prb51}). The CoO
spectrum shows the strongest resemblance to the \TiCoO\ spectrum,
arguing that the valence state of the Co cations in \TiCoO\ is the
divalent high-spin ground state seen in CoO \cite{vanElp1991:prb44}.
This observation confirms the conclusions drawn from comparison of the
rutile \TiCoO\ spectrum with cluster model calculations.

It is worth noting that while general agreement exists regarding the
Co$^{2+}$ oxidation state in both rutile and anatase \TiCoO, there is
widespread disagreement over whether the spin state is high
\cite{Kim2003:prl90,Park2002:prb65,Weng2004:prb69} or low
\cite{Matsumoto2001:jjap40,Matsumoto2001:sci291,Yao2004:jap95}. Most
conclusions of a low spin state have been drawn from the low average
magnetic moment observed in magnetization measurements. It may be
difficult, however, to deduce the spin state from the magnetization
due to the effects of disorder. For example, it has been suggested
that interstitially substituted Co will strongly suppress the spin
moment of the Co ions on the Ti site and thus the average Co magnetic
moment \cite{Geng2003:prb68}.

\begin{figure}
    \centreline{\includegraphics[scale=0.676]{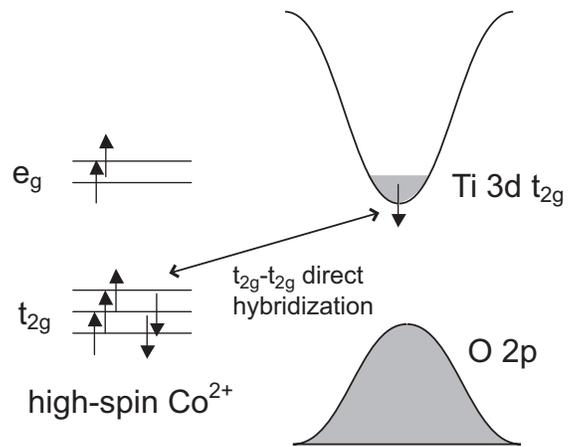}}
    \caption{Schematic band diagram showing the high spin Co~$^{2+}$ state 
             and the resulting strong $t_{2g}$-$t_{2g}$ coupling
             between the Ti~$3d$ and Co~$3d$ $t_{2g}$ states.}
    \label{fig:bandDiagram}
\end{figure}

The high spin Co$^{2+}$ state has a partially unoccupied $t_{2g}$
orbital. This unoccupied Co~$3d$ $t_{2g}$ orbital is expected to
strongly hybridize with the Ti~$3d$ $t_{2g}$ orbital, permitting
direct $t_{2g}$-$t_{2g}$ hopping as shown in
Fig.~\ref{fig:bandDiagram}. Such direct hopping should enhance the
exchange coupling between the Co ions, increasing the magnetic moment
\cite{Furdyna1982:jap53}. Indeed, the rutile phase shows a higher
average magnetic moment in magnetisation measurements
\cite{Toyosaki2004:NatMat3} compared to the anatase phase
\cite{Matsumoto2001:sci291}, consistent with this expectation.

\begin{figure}
    \centreline{\includegraphics{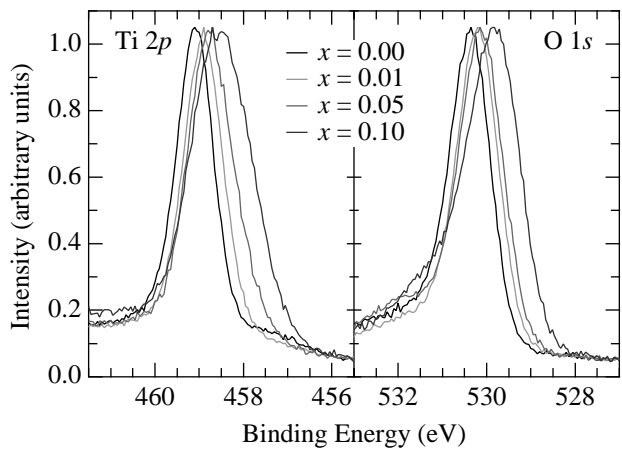}}
    \caption{Ti~$2p$ and O~$1s$ XPS core level spectra as a function
             of cobalt doping showing the chemical potential shift.
             The energy shift due to band bending at the surface is
             removed by laser illumination.}
    \label{fig:Ti2p_O1s_xDep}
\end{figure}

Figure~\ref{fig:Ti2p_O1s_xDep} shows Ti~$2p$ and O~$1s$ core level
spectra for all \TiCoO\ films studied ($x =0$, 0.01, 0.05 and 0.10).
To remove the effect of band bending at the surface, the spectra were
measured under laser irradiation at 3.5~eV. Photoexcitation is known
to induce a surface photovoltage which cancels band bending at the
surface. Consistent with this we observed a small increase in binding
energy with photoexcitation in all samples, including pure \TiO,
signalling the removal of this extrinsic binding energy shift. The
pure \TiO\ film shows a sharp, symmetric peak at a binding energy just
under 530.4~eV, while the $x =0.01$, 0.05 and 0.10 core level peaks
are somewhat broadened (particularly the $x = 0.10$ sample) and are
shifted by 0.13, 0.18 and 0.50~eV to lower binding energy. This
binding energy shift is attributed to a chemical potential shift due
to the exchange splitting of the conduction band
\cite{Toyosaki2004:NatMat3}. The implications of this binding energy
shift---the degeneracy of the Co impurity acceptor level and the conduction
band minimum---were discussed earlier.

In conclusion, we have studied the electronic structure of \TiCoO\
($x=0.01$, 0.05 and 0.10) using x-ray photoemission spectroscopy. The
electronic state of Co in rutile \TiCoO\ is high spin Co$^{2+}$,
indicating substitution of the Co ions on the Ti sites. There is no
discernible metallic Co signal in the XPS spectra. The valence band
spectra show the evolution of a broad impurity band derived from the
Co~$3d$ orbitals located 2~eV below the Fermi level. A cluster model
analysis of the Co~$2p$ core-level spectrum gives a charge transfer
gap of 4~eV, while the magnitude of the exchange splitting of the
conduction band is estimated to be as large as 0.5~eV from the
residual Ti~$2p$ and O~$1s$ core level binding energy shift relative
to pure \TiO. Taking into consideration the charge transfer energy and
the binding energy shift due to the exchange splitting, the acceptor
level of the Co impurity is found to be almost degenerate in energy
with the conduction band minimum. The degeneracy and direct
$t_{2g}$-$t_{2g}$ hopping, arising from hybridization between carriers
in the Ti~$3d$ $t_{2g}$ band and the $t_{2g}$ states of the high-spin
Co$^{2+}$, are expected to enhance the exchange coupling between the
carriers and the Co~$3d$ spins and give the high Curie temperature
observed in \TiCoO. These results are consistent with the view of
rutile \TiCoO\ as a diluted magnetic (wide-gap) semiconductor.

\bibliography{Journal_Abbreviations,TiO2_Papers}

\begin{thebibliography}{22}
\expandafter\ifx\csname natexlab\endcsname\relax\def\natexlab#1{#1}\fi
\expandafter\ifx\csname bibnamefont\endcsname\relax
  \def\bibnamefont#1{#1}\fi
\expandafter\ifx\csname bibfnamefont\endcsname\relax
  \def\bibfnamefont#1{#1}\fi
\expandafter\ifx\csname citenamefont\endcsname\relax
  \def\citenamefont#1{#1}\fi
\expandafter\ifx\csname url\endcsname\relax
  \def\url#1{\texttt{#1}}\fi
\expandafter\ifx\csname urlprefix\endcsname\relax\def\urlprefix{URL }\fi
\providecommand{\bibinfo}[2]{#2}
\providecommand{\eprint}[2][]{\url{#2}}

\bibitem[{\citenamefont{Matsumoto
  et~al.}(2001{\natexlab{a}})\citenamefont{Matsumoto, Murakami, Shono,
  Hasegawa, Fukumura, Kawasaki, Ahmet, Chikyow, Koshihara, and
  Koinuma}}]{Matsumoto2001:sci291}
\bibinfo{author}{\bibfnamefont{Y.}~\bibnamefont{Matsumoto}},
  \bibinfo{author}{\bibfnamefont{M.}~\bibnamefont{Murakami}},
  \bibinfo{author}{\bibfnamefont{T.}~\bibnamefont{Shono}},
  \bibinfo{author}{\bibfnamefont{T.}~\bibnamefont{Hasegawa}},
  \bibinfo{author}{\bibfnamefont{T.}~\bibnamefont{Fukumura}},
  \bibinfo{author}{\bibfnamefont{M.}~\bibnamefont{Kawasaki}},
  \bibinfo{author}{\bibfnamefont{P.}~\bibnamefont{Ahmet}},
  \bibinfo{author}{\bibfnamefont{T.}~\bibnamefont{Chikyow}},
  \bibinfo{author}{\bibfnamefont{S.}~\bibnamefont{Koshihara}},
  \bibnamefont{and} \bibinfo{author}{\bibfnamefont{H.}~\bibnamefont{Koinuma}},
  \bibinfo{journal}{Science} \textbf{\bibinfo{volume}{291}},
  \bibinfo{pages}{854} (\bibinfo{year}{2001}{\natexlab{a}}).

\bibitem[{\citenamefont{Matsumoto
  et~al.}(2001{\natexlab{b}})\citenamefont{Matsumoto, Takahashi, Murakami,
  Koida, Fan, Hasegawa, Fukumura, Kawasaki, Koshihara, and
  Koinuma}}]{Matsumoto2001:jjap40}
\bibinfo{author}{\bibfnamefont{Y.}~\bibnamefont{Matsumoto}},
  \bibinfo{author}{\bibfnamefont{R.}~\bibnamefont{Takahashi}},
  \bibinfo{author}{\bibfnamefont{M.}~\bibnamefont{Murakami}},
  \bibinfo{author}{\bibfnamefont{T.}~\bibnamefont{Koida}},
  \bibinfo{author}{\bibfnamefont{X.-J.} \bibnamefont{Fan}},
  \bibinfo{author}{\bibfnamefont{T.}~\bibnamefont{Hasegawa}},
  \bibinfo{author}{\bibfnamefont{T.}~\bibnamefont{Fukumura}},
  \bibinfo{author}{\bibfnamefont{M.}~\bibnamefont{Kawasaki}},
  \bibinfo{author}{\bibfnamefont{S.}~\bibnamefont{Koshihara}},
  \bibnamefont{and} \bibinfo{author}{\bibfnamefont{H.}~\bibnamefont{Koinuma}},
  \bibinfo{journal}{Jpn.\ J.\ Appl.\ Phys.} \textbf{\bibinfo{volume}{40}},
  \bibinfo{pages}{L1204} (\bibinfo{year}{2001}{\natexlab{b}}).

\bibitem[{\citenamefont{Toyosaki et~al.}(2004)\citenamefont{Toyosaki, Fukumura,
  Yamada, Nakajima, Chikyow, Hasegawa, Koinuma, and
  Kawasaki}}]{Toyosaki2004:NatMat3}
\bibinfo{author}{\bibfnamefont{H.}~\bibnamefont{Toyosaki}},
  \bibinfo{author}{\bibfnamefont{T.}~\bibnamefont{Fukumura}},
  \bibinfo{author}{\bibfnamefont{Y.}~\bibnamefont{Yamada}},
  \bibinfo{author}{\bibfnamefont{K.}~\bibnamefont{Nakajima}},
  \bibinfo{author}{\bibfnamefont{T.}~\bibnamefont{Chikyow}},
  \bibinfo{author}{\bibfnamefont{T.}~\bibnamefont{Hasegawa}},
  \bibinfo{author}{\bibfnamefont{H.}~\bibnamefont{Koinuma}}, \bibnamefont{and}
  \bibinfo{author}{\bibfnamefont{M.}~\bibnamefont{Kawasaki}},
  \bibinfo{journal}{Nature Materials} \textbf{\bibinfo{volume}{3}},
  \bibinfo{pages}{221} (\bibinfo{year}{2004}).

\bibitem[{\citenamefont{Park et~al.}(2002)\citenamefont{Park, Kwon, and
  Min}}]{Park2002:prb65}
\bibinfo{author}{\bibfnamefont{M.~S.} \bibnamefont{Park}},
  \bibinfo{author}{\bibfnamefont{S.~K.} \bibnamefont{Kwon}}, \bibnamefont{and}
  \bibinfo{author}{\bibfnamefont{B.~I.} \bibnamefont{Min}},
  \bibinfo{journal}{Phys.\ Rev.\ B} \textbf{\bibinfo{volume}{65}},
  \bibinfo{pages}{161201(R)} (\bibinfo{year}{2002}).

\bibitem[{\citenamefont{Geng and Kim}(2003)}]{Geng2003:prb68}
\bibinfo{author}{\bibfnamefont{W.~T.} \bibnamefont{Geng}} \bibnamefont{and}
  \bibinfo{author}{\bibfnamefont{K.~S.} \bibnamefont{Kim}},
  \bibinfo{journal}{Phys.\ Rev.\ B} \textbf{\bibinfo{volume}{68}},
  \bibinfo{pages}{125203} (\bibinfo{year}{2003}).

\bibitem[{\citenamefont{Kim et~al.}(2003)\citenamefont{Kim, Park, Park, Noh,
  Oh, Yang, Kim, Bu, Noh, Lin et~al.}}]{Kim2003:prl90}
\bibinfo{author}{\bibfnamefont{J.-Y.} \bibnamefont{Kim}},
  \bibinfo{author}{\bibfnamefont{J.-H.} \bibnamefont{Park}},
  \bibinfo{author}{\bibfnamefont{B.-G.} \bibnamefont{Park}},
  \bibinfo{author}{\bibfnamefont{H.-J.} \bibnamefont{Noh}},
  \bibinfo{author}{\bibfnamefont{S.-J.} \bibnamefont{Oh}},
  \bibinfo{author}{\bibfnamefont{J.~S.} \bibnamefont{Yang}},
  \bibinfo{author}{\bibfnamefont{D.-H.} \bibnamefont{Kim}},
  \bibinfo{author}{\bibfnamefont{S.~D.} \bibnamefont{Bu}},
  \bibinfo{author}{\bibfnamefont{T.-W.} \bibnamefont{Noh}},
  \bibinfo{author}{\bibfnamefont{H.-J.} \bibnamefont{Lin}},
  \bibnamefont{et~al.}, \bibinfo{journal}{Phys.\ Rev.\ Lett.}
  \textbf{\bibinfo{volume}{90}}, \bibinfo{pages}{017401}
  (\bibinfo{year}{2003}).

\bibitem[{\citenamefont{Shinde et~al.}(2003)\citenamefont{Shinde, Ogale, Sarma,
  Simpson, Drew, Lofland, Lanci, Buban, Browning, Kulkarni
  et~al.}}]{Shinde2003:prb67}
\bibinfo{author}{\bibfnamefont{S.~R.} \bibnamefont{Shinde}},
  \bibinfo{author}{\bibfnamefont{S.~B.} \bibnamefont{Ogale}},
  \bibinfo{author}{\bibfnamefont{S.~D.} \bibnamefont{Sarma}},
  \bibinfo{author}{\bibfnamefont{J.~R.} \bibnamefont{Simpson}},
  \bibinfo{author}{\bibfnamefont{H.~D.} \bibnamefont{Drew}},
  \bibinfo{author}{\bibfnamefont{S.~E.} \bibnamefont{Lofland}},
  \bibinfo{author}{\bibfnamefont{C.}~\bibnamefont{Lanci}},
  \bibinfo{author}{\bibfnamefont{J.~P.} \bibnamefont{Buban}},
  \bibinfo{author}{\bibfnamefont{N.~D.} \bibnamefont{Browning}},
  \bibinfo{author}{\bibfnamefont{V.~N.} \bibnamefont{Kulkarni}},
  \bibnamefont{et~al.}, \bibinfo{journal}{Phys.\ Rev.\ B}
  \textbf{\bibinfo{volume}{67}}, \bibinfo{pages}{115211}
  (\bibinfo{year}{2003}).

\bibitem[{\citenamefont{Shinde et~al.}(2004)\citenamefont{Shinde, Ogale,
  Higgins, Zheng, Millis, Kulkarni, Ramesh, Greene, and
  Venkatesan}}]{Shinde2004:prl92}
\bibinfo{author}{\bibfnamefont{S.~R.} \bibnamefont{Shinde}},
  \bibinfo{author}{\bibfnamefont{S.~B.} \bibnamefont{Ogale}},
  \bibinfo{author}{\bibfnamefont{J.~S.} \bibnamefont{Higgins}},
  \bibinfo{author}{\bibfnamefont{H.}~\bibnamefont{Zheng}},
  \bibinfo{author}{\bibfnamefont{A.~J.} \bibnamefont{Millis}},
  \bibinfo{author}{\bibfnamefont{V.~N.} \bibnamefont{Kulkarni}},
  \bibinfo{author}{\bibfnamefont{R.}~\bibnamefont{Ramesh}},
  \bibinfo{author}{\bibfnamefont{R.~L.} \bibnamefont{Greene}},
  \bibnamefont{and}
  \bibinfo{author}{\bibfnamefont{T.}~\bibnamefont{Venkatesan}},
  \bibinfo{journal}{Phys.\ Rev.\ Lett.} \textbf{\bibinfo{volume}{92}},
  \bibinfo{pages}{166601} (\bibinfo{year}{2004}).

\bibitem[{\citenamefont{Chiba et~al.}(2003)\citenamefont{Chiba, Takamura,
  Matsukura, and Ohno}}]{Chiba2003:apl82}
\bibinfo{author}{\bibfnamefont{D.}~\bibnamefont{Chiba}},
  \bibinfo{author}{\bibfnamefont{K.}~\bibnamefont{Takamura}},
  \bibinfo{author}{\bibfnamefont{F.}~\bibnamefont{Matsukura}},
  \bibnamefont{and} \bibinfo{author}{\bibfnamefont{H.}~\bibnamefont{Ohno}},
  \bibinfo{journal}{Appl. Phys. Lett.} \textbf{\bibinfo{volume}{82}},
  \bibinfo{pages}{3020} (\bibinfo{year}{2003}).

\bibitem[{\citenamefont{Woicik et~al.}(2002)\citenamefont{Woicik, Nelson,
  Kronik, Jain, Chelikowsky, Heskett, Berman, and Herman}}]{Woicik2002:prl89}
\bibinfo{author}{\bibfnamefont{J.~C.} \bibnamefont{Woicik}},
  \bibinfo{author}{\bibfnamefont{E.~J.} \bibnamefont{Nelson}},
  \bibinfo{author}{\bibfnamefont{L.}~\bibnamefont{Kronik}},
  \bibinfo{author}{\bibfnamefont{M.}~\bibnamefont{Jain}},
  \bibinfo{author}{\bibfnamefont{J.~R.} \bibnamefont{Chelikowsky}},
  \bibinfo{author}{\bibfnamefont{D.}~\bibnamefont{Heskett}},
  \bibinfo{author}{\bibfnamefont{L.~E.} \bibnamefont{Berman}},
  \bibnamefont{and} \bibinfo{author}{\bibfnamefont{G.~S.}
  \bibnamefont{Herman}}, \bibinfo{journal}{Phys.\ Rev.\ Lett.}
  \textbf{\bibinfo{volume}{89}}, \bibinfo{pages}{077401}
  (\bibinfo{year}{2002}).

\bibitem[{\citenamefont{Moulder et~al.}(1992)\citenamefont{Moulder, Stickle,
  Sobol, and Bomben}}]{XPSHandbook1992}
\bibinfo{author}{\bibfnamefont{J.~F.} \bibnamefont{Moulder}},
  \bibinfo{author}{\bibfnamefont{W.~F.} \bibnamefont{Stickle}},
  \bibinfo{author}{\bibfnamefont{P.~E.} \bibnamefont{Sobol}}, \bibnamefont{and}
  \bibinfo{author}{\bibfnamefont{K.~D.} \bibnamefont{Bomben}},
  \emph{\bibinfo{title}{Handbook of X-ray Photoelectron Spectroscopy}}
  (\bibinfo{publisher}{Physical Electronics Division, Perkin-Elmer Corporation,
  Eden Prairie, MN}, \bibinfo{year}{1992}).

\bibitem[{\citenamefont{Cattaruzza et~al.}(1998)\citenamefont{Cattaruzza,
  Gonella, Mattei, , Mazzoldi, Gatteschi, Sangregorio, Falconieri, Salvetti,
  and Battaglin}}]{Cattaruzza1998:apl73}
\bibinfo{author}{\bibfnamefont{E.}~\bibnamefont{Cattaruzza}},
  \bibinfo{author}{\bibfnamefont{F.}~\bibnamefont{Gonella}},
  \bibinfo{author}{\bibfnamefont{G.}~\bibnamefont{Mattei}}, ,
  \bibinfo{author}{\bibfnamefont{P.}~\bibnamefont{Mazzoldi}},
  \bibinfo{author}{\bibfnamefont{D.}~\bibnamefont{Gatteschi}},
  \bibinfo{author}{\bibfnamefont{C.}~\bibnamefont{Sangregorio}},
  \bibinfo{author}{\bibfnamefont{M.}~\bibnamefont{Falconieri}},
  \bibinfo{author}{\bibfnamefont{G.}~\bibnamefont{Salvetti}}, \bibnamefont{and}
  \bibinfo{author}{\bibfnamefont{G.}~\bibnamefont{Battaglin}},
  \bibinfo{journal}{Applied Physics Letters} \textbf{\bibinfo{volume}{73}},
  \bibinfo{pages}{1176} (\bibinfo{year}{1998}).

\bibitem[{\citenamefont{Wang et~al.}(2004)\citenamefont{Wang, Jiang, and
  Meletis}}]{Wang2004:jap95}
\bibinfo{author}{\bibfnamefont{F.~L.} \bibnamefont{Wang}},
  \bibinfo{author}{\bibfnamefont{J.~C.} \bibnamefont{Jiang}}, \bibnamefont{and}
  \bibinfo{author}{\bibfnamefont{E.~I.} \bibnamefont{Meletis}},
  \bibinfo{journal}{Journal of Applied Physics} \textbf{\bibinfo{volume}{95}},
  \bibinfo{pages}{5069} (\bibinfo{year}{2004}).

\bibitem[{End()}]{Endnote_condmat0507193}
\bibinfo{note}{Taking the average moment of metallic Co clusters to be the bulk
  value of 1.73~$\mu_{B}/$Co.}

\bibitem[{\citenamefont{Furdyna}(1982)}]{Furdyna1982:jap53}
\bibinfo{author}{\bibfnamefont{J.~K.} \bibnamefont{Furdyna}},
  \bibinfo{journal}{J. Appl. Phys.} \textbf{\bibinfo{volume}{53}},
  \bibinfo{pages}{7637} (\bibinfo{year}{1982}).

\bibitem[{\citenamefont{van Elp et~al.}(1991)\citenamefont{van Elp, Wieland,
  Eskes, Kuiper, Sawatzky, de~Groot, and Turner}}]{vanElp1991:prb44}
\bibinfo{author}{\bibfnamefont{J.}~\bibnamefont{van Elp}},
  \bibinfo{author}{\bibfnamefont{J.~L.} \bibnamefont{Wieland}},
  \bibinfo{author}{\bibfnamefont{H.}~\bibnamefont{Eskes}},
  \bibinfo{author}{\bibfnamefont{P.}~\bibnamefont{Kuiper}},
  \bibinfo{author}{\bibfnamefont{G.~A.} \bibnamefont{Sawatzky}},
  \bibinfo{author}{\bibfnamefont{F.~M.~F.} \bibnamefont{de~Groot}},
  \bibnamefont{and} \bibinfo{author}{\bibfnamefont{T.~S.}
  \bibnamefont{Turner}}, \bibinfo{journal}{Phys.\ Rev.\ B}
  \textbf{\bibinfo{volume}{44}}, \bibinfo{pages}{6090} (\bibinfo{year}{1991}).

\bibitem[{\citenamefont{Saitoh et~al.}(1997)\citenamefont{Saitoh, Mizokawa,
  Fujimori, Abbate, Takeda, and Takano}}]{Saitoh1997:prb56}
\bibinfo{author}{\bibfnamefont{T.}~\bibnamefont{Saitoh}},
  \bibinfo{author}{\bibfnamefont{T.}~\bibnamefont{Mizokawa}},
  \bibinfo{author}{\bibfnamefont{A.}~\bibnamefont{Fujimori}},
  \bibinfo{author}{\bibfnamefont{M.}~\bibnamefont{Abbate}},
  \bibinfo{author}{\bibfnamefont{Y.}~\bibnamefont{Takeda}}, \bibnamefont{and}
  \bibinfo{author}{\bibfnamefont{M.}~\bibnamefont{Takano}},
  \bibinfo{journal}{Phys.\ Rev.\ B} \textbf{\bibinfo{volume}{56}},
  \bibinfo{pages}{1290} (\bibinfo{year}{1997}).

\bibitem[{\citenamefont{Bocquet et~al.}(1992)\citenamefont{Bocquet, Mizokawa,
  Saitoh, Namatame, and Fujimori}}]{Bocquet1992:prb46}
\bibinfo{author}{\bibfnamefont{A.~E.} \bibnamefont{Bocquet}},
  \bibinfo{author}{\bibfnamefont{T.}~\bibnamefont{Mizokawa}},
  \bibinfo{author}{\bibfnamefont{T.}~\bibnamefont{Saitoh}},
  \bibinfo{author}{\bibfnamefont{H.}~\bibnamefont{Namatame}}, \bibnamefont{and}
  \bibinfo{author}{\bibfnamefont{A.}~\bibnamefont{Fujimori}},
  \bibinfo{journal}{Phys.\ Rev.\ B} \textbf{\bibinfo{volume}{46}},
  \bibinfo{pages}{3771} (\bibinfo{year}{1992}).

\bibitem[{\citenamefont{Potze et~al.}(1995)\citenamefont{Potze, Sawatzky, and
  Abbate}}]{Potze1995:prb51}
\bibinfo{author}{\bibfnamefont{R.~H.} \bibnamefont{Potze}},
  \bibinfo{author}{\bibfnamefont{G.~A.} \bibnamefont{Sawatzky}},
  \bibnamefont{and} \bibinfo{author}{\bibfnamefont{M.}~\bibnamefont{Abbate}},
  \bibinfo{journal}{Phys.\ Rev.\ B} \textbf{\bibinfo{volume}{51}},
  \bibinfo{pages}{11501} (\bibinfo{year}{1995}).

\bibitem[{\citenamefont{Weng et~al.}(2004)\citenamefont{Weng, Yang, Dong,
  Mizuseki, Kawasaki, and Kawazoe}}]{Weng2004:prb69}
\bibinfo{author}{\bibfnamefont{H.}~\bibnamefont{Weng}},
  \bibinfo{author}{\bibfnamefont{X.}~\bibnamefont{Yang}},
  \bibinfo{author}{\bibfnamefont{J.}~\bibnamefont{Dong}},
  \bibinfo{author}{\bibfnamefont{H.}~\bibnamefont{Mizuseki}},
  \bibinfo{author}{\bibfnamefont{M.}~\bibnamefont{Kawasaki}}, \bibnamefont{and}
  \bibinfo{author}{\bibfnamefont{Y.}~\bibnamefont{Kawazoe}},
  \bibinfo{journal}{Phys.\ Rev.\ B} \textbf{\bibinfo{volume}{69}},
  \bibinfo{pages}{125219} (\bibinfo{year}{2004}).

\bibitem[{\citenamefont{Yao et~al.}(2004)\citenamefont{Yao, Zhou, Gai, Chong,
  and Wang}}]{Yao2004:jap95}
\bibinfo{author}{\bibfnamefont{X.~F.} \bibnamefont{Yao}},
  \bibinfo{author}{\bibfnamefont{T.~J.} \bibnamefont{Zhou}},
  \bibinfo{author}{\bibfnamefont{Y.~X.} \bibnamefont{Gai}},
  \bibinfo{author}{\bibfnamefont{T.~C.} \bibnamefont{Chong}}, \bibnamefont{and}
  \bibinfo{author}{\bibfnamefont{J.~P.} \bibnamefont{Wang}},
  \bibinfo{journal}{J. Appl. Phys.} \textbf{\bibinfo{volume}{95}},
  \bibinfo{pages}{7375} (\bibinfo{year}{2004}).

\bibitem[{\citenamefont{Galtayries and
  Grimblot}(1999)}]{Galtayries1999:jesrp98}
\bibinfo{author}{\bibfnamefont{A.}~\bibnamefont{Galtayries}} \bibnamefont{and}
  \bibinfo{author}{\bibfnamefont{J.}~\bibnamefont{Grimblot}},
  \bibinfo{journal}{Journal of Electron Spectroscopy and Related Phenomena}
  \textbf{\bibinfo{volume}{98--99}}, \bibinfo{pages}{267}
  (\bibinfo{year}{1999}).

\end{thebibliography}

\end{document}